\documentstyle[12pt]{article}

\begin{document}
\thispagestyle{empty}

\begin{center}
               RUSSIAN GRAVITATIONAL ASSOCIATION\\
               CENTER FOR SURFACE AND VACUUM RESEARCH\\
               DEPARTMENT OF FUNDAMENTAL INTERACTIONS AND METROLOGY
\end{center}
\vskip 4ex
\begin{flushright}                              RGA-CSVR-009/94\\
                                                gr-qc/9407019
\end{flushright}
\vskip 45mm

\begin{center}
{\bf
MULTIDIMENSIONAL COSMOLOGY WITH\\ MULTICOMPONENT PERFECT FLUID
AND TODA LATTICES}
 \vskip 5mm
{\bf  V.R. Gavrilov, V. D. Ivashchuk and V. N. Melnikov}\\

\vskip 5mm
     {\em Centre for Surface and Vacuum Research,\\
     8 Kravchenko str., Moscow, 117331, Russia}\\
     e-mail: mel@cvsi.uucp.free.net\\
\vskip 10mm
\end{center}
\begin{abstract}
The integration procedure  for multidimensional cosmological models
with multicomponent  perfect  fluid    in
spaces   of   constant   curvature   is   developed.   Reduction  of
pseudo-Euclidean Toda-like systems to the Euclidean  ones  is  done.
Some known solutions are singled out from those obtained. The existence
of the wormholes is proved.
\end{abstract}

\vskip 10mm

\begin{center}


\vskip 20mm

             Moscow 1994
\end{center}
\pagebreak

\section{Introduction}
\par
We consider dynamical systems with $n\geq 2$  degrees of freedom
described by the Lagrangian
\begin{equation} L=\frac{1}{2} \sum_{i,j=1}^{n}
\eta_{ij} \dot{x}^{i} \dot{x}^{j} - \sum_{s=1}^{m}a^{(s)} \exp[ \sum_{i=1}^{n}
b_i^sx^i], \ \ m\geq 2.
\end{equation}
A lot of systems in gravitation [10,11] and as well in multidimensional
cosmology [12-34] reduce to the systems with such a Lagrangian.

Without loss of generality it can be assumed that matrix $(\eta_{ij})$ is
diagonalized and $\eta_{ii}=\pm 1$ for $i=1,\ldots ,m$. Such  system is an
algebraic generalization of  a  well-known   Toda lattice [1]
suggested by Bogoyavlensky [2]. We say that it is an Euclidean Toda-like
system, if bilinear form of kinetic energy is positively definite, i.e.
$\eta_{ij}=\delta_{ij}$. Nearly nothing is known about  Euclidean
Toda-like systems with arbitrary sets of vectors $b_{1},\ldots ,b_{m}$,
where $b_{s}=(b_{1}^{s},\ldots ,b_{n}^{s})$ for $s=1,\ldots ,m$. But, if they
form the set of admissible roots of a simple Lie algebra, then the system
is completely integrable and possesses a Lax representation.  Remind, that
the set of roots $\alpha_{1},\ldots,\alpha_{m}$ is called admissible [2],
provided vectors $\alpha_{r} - \alpha_{s}$ are not roots for all
$r,s=1,\ldots ,m$. Each simple Lie algebra  possesses the following  set
of admissible roots
\begin{equation}
\omega_{1},\ldots ,\omega_{n},-\Omega
\end{equation}
where $\omega_{1},\ldots ,\omega_{n}$ are simple roots and $\Omega $ is
maximal root [3] (usually $\Omega=\omega_{1}+\ldots +\omega_{n}$). Any subset
of the set (1.2) is also admissible.

If the maximal root holds in the set (1.2), then generalized periodic Toda
lattices arise. The different $L-A$ pairs for them were found by Bogoyavlensky
[2].  There were also presented the Hamiltonians of this systems connected
with simple Lie algebras.

The further progress in this field was
attained by a number of authors (see, for example, [4-8] and refs. therein).
In ref. [6] Adler and van Moerbeke established a criterion of algebraic
complete integrability for Euclidean Toda-like systems. (This criterion was
formally applied to multidimensional vacuum cosmology with $n$ Einstein
spaces in [30].) The explicit integration of the equations of motion
for the generalized open Toda lattices (in this case the maximal root is
thrown away) was developed by Olshanetsky and Perelomov [4] and Kostant [5].
(See also [7].)

Here we are interested in the problem of integrability of the Toda-like
systems with the indefinite bilinear form of the kinetic energy.  Let us call
such systems  pseudo-Euclidean Toda-like  systems. To our knowledge, this
problem has not been discussed intensively in the mathematical literature
before.  The reason, as it seems to us, consists in the following.  If one try
to connect a pseudo-Euclidean Toda-like system by the known manner with
simple Lie algebra   it reduces to an Euclidean system for the part of
coordinates (see Sect. IV).  Nevertheless, integrable pseudo-Euclidean
Toda-like systems and search for their solutions in explicit form evoke a
special interest, because such systems arise in cosmology. For instance,
4-dimensional vacuum homogeneous cosmological model of Bianchi IX-type is
described by the Lagrangian (1.1) with $(\eta_{ij})=$diag$(-1, +1, +1)$
[2]. (In [9] it was shown, that this model has a rather rich mathematical
structure.)

So, in the present paper we study integrable pseudo-Euclidean Toda-like
systems appearing  in multidimensional cosmology [12-34].  This direction in
the modern theoretical physics has appeared within the new paradigm  based
on unified  theories and hypothesis  of additional space-time dimensions.
According to this  hypothesis  the physical space-time manifold has the
topology $M^{4}\times B$, where $M^{4}$ is a 4-dimensional manifold,  and B
is a so-called internal  space  (or spaces).  Nonobservability  of additional
dimensions is attained in multidimensional cosmology  by spontaneous or
dynamical compactification of internal spaces to the Planck scale
($10^{-33}$ cm.). Integrable cosmological models are of great interest,
because the   exact solutions allow to study dynamical
properties of the model,  in particular compactification of internal spaces,
in detail.

In the Sect. II, as in [34], we consider  the cosmological  model
where multidimensional space-time manifold $M$ is a direct product
of the time axis R and of the $n$ Einstein spaces $M_{1},\ldots,M_{n}$.
We remind, that any manifold of constant curvature is the Einstein
one.  It is shown that Einstein  equations  for  the scale factors with a
source in the multicomponent perfect fluid form correspond to the
Lagrangian (1.1) with $(\eta_{ij})=$diag$(-1, +1,\ldots , +1)$. We
develop  the integration procedure to the case of  an orthogonal
set of vectors $b_{1},\ldots ,b_{m}$ in Sect. III. Sect. IV is
devoted to the reduction of pseudo-Euclidean Toda-like system to
the Euclidean one for a part of coordinates. This reduction allows us to
obtain the class of the exact solutions for  some nonorthogonal
sets of the vectors $b_{1},\ldots ,b_{m}$.  We present the exact
solution in the simplest case, when the reducible pseudo-Euclidean
system is connected with the Lie algebra $A_{2}$.  Discussion of
results is presented is Sect. V. We single out some interesting
solutions, in particular, Euclidean wormholes.

We denote by $n$ the number of  Einstein spaces  and by $m$ the
number of the matter components. Indexes $i$ and $j$ run from
1 to $n$. Index $s$ runs from 1 to $m$.

\section{The model}

We consider  a cosmological model describing
the evolution of $n\geq 2$ Einstein spaces in the presence of $m$-component
perfect-fluid matter [34]. The metric of the model
\begin{equation} g=-\exp[2{\gamma}(t)]dt \otimes dt +
\sum_{i=1}^{n}\exp[2{x^{i}}(t)]g^{(i)},
\end{equation}
is defined on the  manifold
 \begin{equation}
M = R \times M_{1} \times \ldots \times M_{n},
\end{equation}
where the manifold $M_{i}$ with the metric $g^{(i)}$ is an
Einstein space of dimension $N_{i}$, i.e.
\begin{equation}
{R_{m_{i}n_{i}}}[g^{(i)}] = \lambda^{i} g^{(i)}_{m_{i}n_{i}}, \ \
m_{i},n_{i}=1,\ldots ,N_{i}.
\end{equation}

The energy-momentum tensor is taken in the following form
\begin{eqnarray}
&& T^{M}_{N} = \sum_{\alpha =1}^{m} T^{M (\alpha)}_{N}, \\
&&(T^{M (\alpha)}_{N})={\rm diag}(-{\rho^{(\alpha)}}(t),
 {p_{1}^{(\alpha)}}(t) \delta^{m_{1}}_{k_{1}},
\ldots , {p^{(\alpha)}_{n}}(t) \delta^{m_{n}}_{k_{n}}),
\end{eqnarray}
with the conservation law constraints imposed:
\begin{equation}
\bigtriangledown_{M} T^{M (\beta)}_{N}=0 \end{equation}
$\beta =1, \ldots ,m-1$.
The Einstein equations
\begin{equation}
R^{M}_{N}-\frac{1}{2}\delta^{M}_{N}R=\kappa^{2}T^{M}_{N}
\end{equation}
($\kappa^{2}$ is gravitational constant) imply
$\bigtriangledown_{M} T^{M}_{N}=0$ and consequently
$\bigtriangledown_{M} T^{M (m)}_{N}=0$.

We suppose that for any $\alpha$-th component of matter the
pressures in all spaces are proportional to the density
\begin{equation}
{p_{i}^{(\alpha)}}(t) = (1-{h_{i}^{(\alpha)}}){\rho^{(\alpha)}}(t),
\end{equation}
where \begin{equation}
{h_{i}^{(\alpha)}} ={\rm const}.
\end{equation}

The non-zero components of the Ricci-tensor for the metric
(2.1) are the following
\begin{equation}
R_{00}=- \sum_{i=1}^{n} N_{i}[ \ddot{x}^{i} - \dot{\gamma} \dot{x}^{i}
+ (\dot{x}^{i})^{2}], \end{equation} \begin{equation}
R_{m_{i}n_{i}}=g_{m_{i}n_{i}}^{(i)} [\lambda^{i} +\exp[2x^{i}-2\gamma]
(\ddot{x}^{i}+\dot{x}^{i}(\sum_{i=1}^{n}N_{i}\dot{x}^{i}-\dot{\gamma}))],
\end{equation}
$i=1,\ldots,n$.

We put \begin{equation}
\gamma = \gamma_{0} \equiv \sum_{i=1}^{n} N_{i}x^{i} \end{equation}
in (1.1) (the harmonic time is used).

The conservation law constraint (2.6) for $\alpha=1,\ldots,m$ reads
\begin{equation}
\dot{\rho}^{(\alpha)}+
\sum_{i=1}^{n}N_{i}\dot{x}^{i}(\rho^{(\alpha)} + p_{i}^{(\alpha)})=0.
\end{equation}
We impose the conditions of state in the form (2.8), (2.9). Then
eq. (2.13) gives  \begin{equation}
{\rho^{(\alpha)}}(t)=A^{(\alpha)}
\exp[-2\gamma_{0} + \sum_{i=1}^{n} u_{i}^{(\alpha)} x^{i}].
\end{equation}
where $A^{(\alpha)}=$const and
\begin{equation}
u_{i}^{(\alpha)}=N_{i}h_{i}^{(\alpha)}.
\end{equation}

The Einstein eqs. (2.7) may be written in the following manner
\begin{equation} \frac{1}{2}\sum_{i,j=1}^{n}G_{ij}\dot{x}^{i}\dot{x}^{j}+ V=0
, \end{equation} \begin{equation}
\lambda^{i}+ \ddot{x}^{i}\exp[2x^{i}-2\gamma_{0}]=
- \kappa^{2} \sum_{\alpha=1}^{m} u^{i}_{(\alpha)} A^{(\alpha)}
\exp[2x^{i}-2\gamma_{0}+\sum_{j=1}^{n} u_{j}^{(\alpha)}x^{j}].
\end{equation}
Here
\begin{equation}
G_{ij}=N_{i}\delta_{ij}-N_{i}N{j}
\end{equation}
are the components of the minisuperspace metric,
\begin{equation}
V=-\frac{1}{2}\sum_{i=1}^{n}\lambda^{i}N_{i}\exp[-2x^{i}+2\gamma_{0}]+
\kappa^2\sum_{\alpha=1}^{m}A^{(\alpha)}\exp[\sum_{i=1}^{n}u_{i}^{(\alpha)}
x^{i}].
\end{equation}
We denote
\begin{equation}
u_{(\alpha)}^{i}=\sum_{j=1}^{n}G^{ij}u_{j}^{(\alpha)},
\end{equation}
where
\begin{equation}
G^{ij} = \frac{\delta^{ij}}{N_{i}}+\frac{1}{2-D}
\end{equation}
are the components of the matrix inverse to the one $(G_{ij})$
[24].

It is  not  difficult  to  verify  that  eqs.  (2.17)  are
equivalent to the Lagrange-Euler eqs. for the Lagrangian
\begin{equation}
L=\frac{1}{2}\sum_{i,j=1}^{n}G_{ij}\dot{x}^{i}\dot{x}^{j}-V.
\end{equation}
Eq. (2.16) is the zero-energy constraint.

We note, that  in the framework of our model the curvature induced
terms in the  potential  (2.19)  may  be  considered  as  additional
components of   the   perfect   fluid.   The   introduction  of  the
cosmological constant $\Lambda$ into the model is equivalent also to  the
addition of a new component with $u_{i}=2N_{i}$ and $\kappa^{2}A=\Lambda$.

Finally, we present the potential (2.19) modified by introduction of
 $\Lambda$-term in the following form
\begin{eqnarray}
V=&&\sum_{k=1}^{n}(-\frac{1}{2}\lambda^{k}N_{k})\exp[\sum_{i,j=1}^{n}G_{ij}
v_{(k)}^{i}x^{j}]+ \nonumber \\
&&\sum_{\alpha=1}^{m}\kappa^{2}A^{(\alpha)}
\exp[\sum_{i,j=1}^{n}G_{ij}u_{(\alpha)}^{i}x^{j}]+
\Lambda\exp[\sum_{i,j=1}^{n}G_{ij}u^{i}x^{j}],
\end{eqnarray}
where we denote:
\begin{eqnarray}
&&v_{(k)}^{i}=\sum_{j=1}^{n}G^{ij}v_{j}^{(k)}=
-2\frac{\delta_{k}^{i}}{N_{k}},\ \ \
 v_{j}^{(k)}\equiv 2(N_{j}-\delta_{j}^{k}), \\
&&u^{i}=\sum_{j=1}^{n}G^{ij}u_{j}.
\end{eqnarray}

Let $<.,.>$ be  a symmetrical bilinear form defined on n-dimensional
real vector space $R^{n}$  with the components
$G_{ij}=<e_{i},e_{j}>$ in the canonical basis $e_{1},\ldots e_{n}$.
($e_{1}=(1,0,\ldots,0$) etc.)
It was shown [22,24], that the bilinear form $<.,.>$ is pseudo-Euclidean one
with the signature $(-, +, ..., +)$.  Then the Lagrangian
(2.22) may be written as:
\begin{equation}
L=\frac{1}{2}<\dot{x},\dot{x}>-\sum_{\alpha=1}^{m}a^{(\alpha)}
\exp[<b_{\alpha},x>].
\end{equation}
($x=x^{1}e_{1}+\ldots+x^{n}e_{n},\ \ x\in R^{n}$). Here we denoted by $m$
 the total number of components,  including
curvature, perfect fluid and the cosmological term. We note, that for $m=1$
the Lagrangian system (2.26) is always integrable. The exact
solutions were obtained in [34]. (Some special cases were considered
in [31,33].) In the
present paper we consider multicomponent case: $m\geq 2$.

We say that a vector $y\in R^{n}$ is called time-like, space-like
or isotropic, if $<y,y>$  has  negative,  positive  or  null values
correspondingly. Vectors $y$ and $z$ are called orthogonal if
$<y,z>=0$.

\section{Exact solutions for orthogonal sets of vectors}
\par
Let vectors $b_{1},\ldots ,b_{m}$ satisfy the conditions:
1. They are linear independent;
2. $<b_{\alpha},b_{\beta}>=0$ for all $\alpha\neq\beta$, i.e. the set of
vectors is orthogonal.

Then $m\leq n$. It is not difficult to prove

Proposition 1.  The set of vectors $b_{1},\ldots ,b_{n}$ may contain
at most one isotropic vector.

Proposition 2.  The set of vectors $b_{1},\ldots ,b_{n}$
may contain at most one time-like vector and, if it holds the
other vectors must be space-like.

Remark 1. The additional term $a^{(0)}\exp[<b_{0},x>]$
with zero-vector $b_{0}=0$ does not change the equations of
motion, but changes the energy constraint (2.16)
\begin{equation}
\frac{1}{2}<\dot{x},\dot{x}>+
\sum_{\alpha=1}^{m}a^{(\alpha)}\exp[<b_{\alpha},x>]
+a^{(0)}=0.
\end{equation}
It corresponds to the  perfect  fluid  with $h_{i}^{(0)}= 0$ for all
$i=1,\ldots ,n$. Such a perfect fluid is called the stiff or Zeldovich
matter [35]. It may be considered also as minimally  coupled  real
scalar  field [36]. We take into account this additional component
by modification of the energy constraint
\begin{equation}
\frac{1}{2}<\dot{x},\dot{x}>+
\sum_{\alpha=1}^{m}a^{(\alpha)}\exp[<b_{\alpha},x>]=E_{0}.
\end{equation}

These propositions allow to split the class of exact  solutions
under consideration into following subclasses:

A. There are one time-like  vector  and  at  most $(n-1)$  space-like
   vectors.

B. There are at most $(n-1)$ space-like vectors.

C. There  are  one  isotropic  vector  and  at most $(n-2)$ space-like
   vectors (this subclass arises for $n\geq 3$).

To integrate  eqs.  of  motion in all subclasses we consider an
orthonormal basis $e_{1}^{\prime},\ldots ,e_{n}^{\prime}$.
These vectors are such that
\begin{equation}
<e_{i}^{\prime},e_{j}^{\prime}>=\eta_{ij},
\end{equation}
where we denote by $\eta_{ij}$ the components of the matrix
\begin{equation}
(\eta_{ij})={\rm diag}(-1,+1,\ldots ,+1).
\end{equation}
Let us define coordinates of the vectors in this basis by
\begin{equation}
x=X^{1}e_{1}^{\prime}+\ldots+X^{n}e_{n}^{\prime}.
\end{equation}
For these new coordinates we have
\begin{equation}
X^{i}=\eta_{ii}<e_{i}^{\prime},x>,\ \ \
x^{i}=\sum_{k=1}^{n}t_{k}^{i}X^{k},
\end{equation}
where we denoted by $t_{k}^{i}$ the components of a non-degenerate matrix
defined by
\begin{equation}
e_{k}^{\prime}=\sum_{i=1}^{n}t_{k}^{i}e_{i}.
\end{equation}
Components $t_{k}^{i}$ satisfy the relations:
\begin{equation}
\sum_{k,l=1}^{n}G_{kl}t_{i}^{k}t_{j}^{l}=\eta_{ij}.
\end{equation}

Let us try to find exact solutions for subclasses A, B and C.

A. Let $b_{1}$ be a time-like vector. Then $<b_{r},b_{r}>>0$
for $r =2,\ldots,m$ (in this case $m\leq n$). We choose the
orthonormal basis $e_{1}^{\prime},\ldots, e_{n}^{\prime}$ as
\begin{equation}
e_{s}^{\prime}=b_{s}/|<b_{s},b_{s}>|^{1/2},\ \
s=1,\ldots,m.
\end{equation}
Then we have:
\begin{equation}
<b_{s},x>=\eta_{ss}|<b_{s},b_{s}>|^{1/2}X^{s}.
\end{equation}
The Lagrangian  (2.26)  and  the  energy  constraint  (3.2)  for the
coordinates $X^{1},\ldots,X^{n}$ have the form
\begin{eqnarray}
&&L=\frac{1}{2}\sum_{i,j=1}^{n}\eta_{ij}\dot{X}^{i}\dot{X}^{j}-
\sum_{s=1}^{m}a^{(s)}\exp[\eta_{ss}|<b_{s},b_{s}>|^{1/2}X^{s}],\\
&&E_{0}=\frac{1}{2}\sum_{i,j=1}^{n}\eta_{ij}\dot{X}^{i}\dot{X}^{j}+
\sum_{s=1}^{m}a^{(s)}\exp[\eta_{ss}|<b_{s},b_{s}>|^{1/2}X^{s}].
\end{eqnarray}
Lagrangian (3.11) leads to the set of eqs.
\begin{eqnarray}
&&\ddot{X}^{s}=-|<b_{s},b_{s}>|^{1/2}a^{(s)}
\exp[\eta_{ss}|<b_{s},b_{s}>|^{1/2}X^{s}], \\
&&\ddot{X}^{m+1}=\ldots =\ddot{X}^{n}=0,
\end{eqnarray}
which is easily integrable. We get
\begin{eqnarray}
&&X^{s}=-\eta_{ss}|<b_{s},b_{s}>|^{-1/2}\ln[F_{s}^{2}(t-t_{0s})],\\
&&X^{m+1}=p^{m+1}t+q^{m+1},\ldots,X^{n}=p^{n}t+q^{n},
\end{eqnarray}
where we denoted
\begin{eqnarray}
F_{s}(t-t_{0s})&&=\sqrt{|a^{(s)}/E_{s}|}
\cosh[\sqrt{|E_{s}<b_{s},b_{s}>|/2}(t-t_{0s})],\ \
{\rm if}\  \eta_{ss}a^{s}>0,\ \eta_{ss}E_{s}>0,\nonumber\\
&&=\sqrt{|a^{(s)}/E_{s}|}
\sin[\sqrt{|E_{s}<b_{s},b_{s}>|/2}(t-t_{0s})],\ \
{\rm if}\  \eta_{ss}a^{s}<0,\ \eta_{ss}E_{s}<0,\nonumber\\
&&=\sqrt{|a^{(s)}/E_{s}|}
\sinh[\sqrt{|E_{s}<b_{s},b_{s}>|/2}(t-t_{0s})],\ \
{\rm if}\  \eta_{ss}a^{s}<0,\ \eta_{ss}E_{s}>0,\nonumber\\
&&=\sqrt{|<b_{s},b_{s}>a^{(s)}|/2}(t-t_{0s}),\ \
{\rm if}\  \eta_{ss}a^{s}<0,\ E_{s}=0.
\end{eqnarray}
By $t_{0s}$, $E_{0s}\ (s=1,\ldots,m),\ p^{m+1},\ldots,p^{n}$,
$q^{m+1},\ldots,q^{n}$ we denoted the integration constants. Some of
them are not arbitrary and connected by the relation
\begin {equation}
E_{1}+\ldots+E_{m}+\frac{1}{2}(p^{m+1})^{2}+\ldots+
\frac{1}{2}(p^{n})^{2}=E_{0}.
\end{equation}
We have for components $t_{k}^{i}$
\begin{equation}
t_{s}^{i}=b_{s}^{i}/|<b_{s},b_{s}>|^{1/2}.
\end{equation}

It is convenient to present the exact solution in a Kasner-like
form. Kasner-like parameters are defined by
\begin{eqnarray}
&&\alpha^{i}=t_{m+1}^{i}p^{m+1}+\ldots+t_{n}^{i}p^{n},\\
&&\beta^{i}=t_{m+1}^{i}q^{m+1}+\ldots+t_{n}^{i}q^{n}.
\end{eqnarray}
Then for the scale factors of the spaces $M_{i}$  (see (3.6)) we get
\begin{equation}
\exp[x^{i}]=\prod_{s=1}^{m}[F_{s}^{2}(t-t_{0s})]
^{-b_{s}^{i}/<b_{s},b_{s}>}\exp[\alpha^{i}t+\beta^{i}].
\end{equation}
Vectors $\alpha,\ \beta \in R^{n}$, are defined by
\begin{eqnarray}
\alpha&=\alpha^{1}e_{1}+\ldots+\alpha^{n}e_{n},\ \
\beta&=\beta^{1}e_{1}+\ldots+\beta^{n}e_{n}
\end{eqnarray}
satisfy the relations
\begin{eqnarray}
&&<\alpha,\alpha>=2(E_{0}-E_{1}-\ldots-E_{m})\geq 0,\\
&&<\alpha,b_{s}>=<\beta,b_{s}>=0,\ \ s=1,\ldots,m.
\end{eqnarray}
We remind that $<\alpha,\beta>=\sum_{i,j=1}^{n}G_{ij}\alpha^{i}\beta^{j}$.

Remark 2. If $m = n$ then $\alpha=\beta=0$.

Remark 3.  The set of constants $E_{0}$,  $E_{s}$,  $t_{0s}$,
$\alpha^{i}$ and $\beta^{i}$ is the final set. Only $2n$ constants
from them are independent.

Remark 4. The subclass of the solutions may be easily enlarged. It is clear,
that the addition of new component inducing
a vector collinear to one of
$b_{1},\ldots,b_{m}$ leads to the integrable by quadrature model. Let
us take into account the following additional terms in the Lagrangian (2.26)
\begin{equation}
-\sum_{\alpha=1}^{m(\sigma)}a^{(\sigma \alpha)}
\exp[b_{(\sigma \alpha)}<b_{\sigma},x>],
\end{equation}
where $b_{(\sigma \alpha)}=$const$\neq 0$ for
$\alpha=1,\ldots,m(\sigma)$, $1\leq \sigma \leq m$. It is not
difficult to show, that the modification of the exact solution
(3.22) only consists in the replacement of the function
$F_{\sigma}(t-t_{0\sigma})$ by  one $F(t-t_{0\sigma})$, satisfying the
quadrature
\begin{equation} \int dF/\sqrt{E_{\sigma}F^{2}-a^{(\sigma)}-
\sum_{\alpha=1}^{m(\sigma)}a^{(\sigma\alpha)}F^{2(1-b_{(\sigma\alpha)})}}
=<b_{\sigma},b_{\sigma}>(t-t_{0\sigma}).
\end{equation}
The additional components with other numbers $\sigma$
may be taken into account by the same manner.

B. We have the set of $m$ space-like vectors $b_{1},\ldots,b_{m}$
($m\leq n-1$) and the orthonormal basis defined by
\begin{equation}
e_{s+1}^{\prime}=b_{s}/\sqrt{<b_{s},b_{s}>},\ \
s=1,\ldots,m.
\end{equation}
The Lagrangian (2.26) and the energy constraint (3.2) in terms of
$X$-coordinates have the form
\begin{eqnarray}
&&L=\frac{1}{2}\sum_{i,j=1}^{n}\eta_{ij}\dot{X}^{i}\dot{X}^{j}-
\sum_{s=1}^{m}a^{(s)}\exp[\sqrt{<b_{s},b_{s}>}X^{s+1}],\\
&&E_{0}=\frac{1}{2}\sum_{i,j=1}^{n}\eta_{ij}\dot{X}^{i}\dot{X}^{j}+
\sum_{s=1}^{m}a^{(s)}\exp[\sqrt{<b_{s},b_{s}>}X^{s+1}].
\end{eqnarray}
The corresponding eqs. of motion
\begin{eqnarray}
&&\ddot{X}^{1}=\ddot{X}^{m+2}=\ldots =\ddot{X}^{n}=0,\\
&&\ddot{X}^{s+1}=-\sqrt{<b_{s},b_{s}>}a^{(s)}
\exp[\sqrt{<b_{s},b_{s}>}X^{s+1}]
\end{eqnarray}
lead to the solution
\begin{eqnarray}
&&X^{1}=p^{1}t+q^{1},\\
&&X^{s+1}=\frac{-1}{\sqrt{<b_{s},b_{s}>}}\ln[F_{s}^{2}(t-t_{0s})],\\
&&X^{m+2}=p^{m+2}t+q^{m+2},\ldots , X^{n}=p^{n}t+q^{n},
\end{eqnarray}
where functions $F_{s}(t-t_{0s})$ are defined by (3.17) (in this case
all $\eta_{ss}=1$). Some of integration constants in (3.33)-(3.35)
satisfy the relation
\begin{equation}
E_{1}+\ldots+E_{m}-\frac{1}{2}(p^{1})^{2}+
\frac{1}{2}(p^{m+2})^{2}+\ldots+\frac{1}{2}(p^{n})^{2}=E_{0}.
\end{equation}
To present the scale factors in a Kasner-like form we define the
parameters:
\begin{eqnarray}
&&\alpha^{i}=t_{1}^{i}p^{1}+t_{m+2}^{i}p^{m+2}+\ldots+t_{n}^{i}p^{n},\\
&&\beta^{i}=t_{1}^{i}q^{1}+t_{m+1}^{i}q^{m+2}+\ldots+t_{n}^{i}q^{n}.
\end{eqnarray}
Then from (3.6) we obtain the same formula:
\begin{equation}
\exp[x^{i}]=\prod_{s=1}^{m}[F_{s}^{2}(t-t_{0s})]
^{-b_{s}^{i}/<b_{s},b_{s}>}\exp[\alpha^{i}t+\beta^{i}].
\end{equation}
The relations (3.8) lead to the following constraints for the
Kasner-like parameters $\alpha^{i}$ and $\beta^{i}$ :
\begin{eqnarray}
&&<\alpha,\alpha>=2(E_{0}-E_{1}-\ldots-E_{m}),\\
&&<\alpha,b_{s}>=<\beta,b_{s}>=0,\ \ s=1,\ldots,m.
\end{eqnarray}

Remark 5.  If $m=n-1$,  then either $<\alpha,\alpha>< 0$ or
$\alpha=0$; and $\beta$ has  the same properties.

Remark 6. We may also consider the enlargement of this subclass
by the manner described in Remark 4. If we add to the Lagrangian
(2.26) the terms (3.26) for the some $\sigma\leq m$, we should
replace the function $F_{\sigma}(t-t_{0\sigma}$) in eq.
(3.39) by the function $F(t-t_{0\sigma})$, satisfying (3.27).

C. Let $b_{1}$ be an isotropic vector. Then $<b_{r},b_{r}>>0$ for
$r=2,\ldots,m$ (in this case $m\leq n-1$). We choose the orthonormal
basis $e_{1}^{\prime},\ldots,e_{n}^{\prime}$ by
\begin{equation}
e_{r}^{\prime}=b_{r}/\sqrt{<b_{r},b_{r}>},\ \
b_{1}=e_{1}^{\prime}+e_{m+1}^{\prime}.
\end{equation}
Then we get
\begin{eqnarray}
&&L=\frac{1}{2}\sum_{i,j=1}^{n}\eta_{ij}\dot{X}^{i}\dot{X}^{j}-
a^{(1)}\exp[-X^{1}+X^{m+1}]-
\sum_{r=2}^{m}a^{(r)}exp[\sqrt{<b_{r},b_{r}>}X^{r}],\\
&&E_{0}=\frac{1}{2}\sum_{i,j=1}^{n}\eta_{ij}\dot{X}^{i}\dot{X}^{j}+
a^{(1)}\exp[-X^{1}+X^{m+1}]+
\sum_{r=2}^{m}a^{(r)}\exp[\sqrt{<b_{r},b_{r}>}X^{r}],
\end{eqnarray}
The corresponding eqs. of motion have the form
\begin{eqnarray}
&&\ddot{X}^{1}=-a^{(1)}\exp[-X^{1}+X^{m+1}],\\
&&\ddot{X}^{m+1}=-a^{(1)}\exp[-X^{1}+X^{m+1}],\\
&&\ddot{X}^{r}=-\sqrt{<b_{r},b_{r}>}a^{(r)}
\exp[\sqrt{<b_{r},b_{r}>}X^{r}],\\
&&\ddot{X}^{m+2}=\ldots=\ddot{X}^{n}=0.
\end{eqnarray}
To integrate (3.45), (3.46) it is useful to consider the eqs. of
motion for $X^{+}=X^{1}+X^{m+1}$ and $X^{-}=-X^{1}+X^{m+1}$. Then we
get the solution
\begin{eqnarray}
&&X^{1}=\frac{1}{2}(p^{+}-p^{-})t+\frac{1}{2}(q^{+}-q^{-})-2\ln[f(t)],\\
&&X^{m+1}=\frac{1}{2}(p^{+}+p^{-})t+\frac{1}{2}(q^{+}+q^{-})-2\ln[f(t)],\\
&&X^{r}=\frac{-1}{\sqrt{<b_{r},b_{r}>}}\ln[F_{r}^{2}(t-t_{0r})],\\
&&X^{m+2}=p^{m+2}t+q^{m+2},\ldots ,X^{n}=p^{n}t+q^{n},
\end{eqnarray}
Here by $f(t)$ we denoted the function
\begin{eqnarray}
f(t)&&=\exp[\frac{a^{(1)}}{2(p^{-})^{2}}\exp[p^{-}t+q^{-}]],\ \
p^{-}\neq 0,\\
&&=\exp[\frac{a^{(1)}}{4}\exp[q^{-}]t^{2}],\ \ p^{-}=0.
\end{eqnarray}
The integration constants satisfy the relations
\begin{eqnarray}
\frac{1}{2}p^{+}p^{-}+E_{2}+\ldots+E_{m}+\frac{1}{2}(p^{m+2})^{2}+
\ldots+\frac{1}{2}(p^{n})^{2}=E_{0},\ \ p^{-}\neq 0,\\
a^{(1)}\exp[q^{-}] +E_{2}+\ldots+E_{m}+
\frac{1}{2}(p^{m+2})^{2}+\ldots+\frac{1}{2}(p^{n})^{2}=E_{0},\ \ p^{-}=0.
\end{eqnarray}
The Kasner-like parameters are defined by
\begin{eqnarray}
&&\alpha^{i}=\frac{1}{2}t_{1}^{i}(p^{+}-p^{-})+
\frac{1}{2}t_{m+1}^{i}(p^{+}+p^{-})+
t_{m+2}^{i}p^{m+2}+\ldots+t_{n}^{i}p^{n},\\
&&\beta^{i}=\frac{1}{2}t_{1}^{i}(q^{+}-q^{-})+
\frac{1}{2}t_{m+1}^{i}(q^{+}+q^{-})+
t_{m+2}^{i}q^{m+2}+\ldots+t_{n}^{i}q^{n}.
\end{eqnarray}
Then from (3.6) we obtain the scale factors in a Kasner-like form:
\begin{equation}
\exp[x^{i}]=[f(t)]^{-b_{1}^{i}}\prod_{r=2}^{m}[F_{r}^{2}(t-t_{0r})]
^{-b_{r}^{i}/<b_{r},b_{r}>}\exp[\alpha^{i}t+\beta{i}].
\end{equation}
The Kasner-like parameters satisfy
\begin{eqnarray}
<\alpha,\alpha>&&=2(E_{0}-E_{2}-\ldots-E_{m}),\ \  <\alpha,b_{1}>\neq 0\\
&&=(E_{0}-a^{(1)}\exp[<\beta,b_{1}>]-E_{2}-\ldots-E_{m}),\ \
<\alpha,b_{1}>=0,
\end{eqnarray}
\begin{equation}
<\alpha,b_{r}>=<\beta,b_{r}>=0,\ \ r=2,\ldots,m.
\end{equation}

Remark 7. For the parameters $p^{-}$ and $q^{-}$ we get:
\begin{equation}
p^{-}=<\alpha,b_{1}>,\ \ \ q^{-}=<\beta,b_{1}>.
\end{equation}

Remark 8. If $m=n-1$ and $<\alpha,b_{1}>=0$, then $<\alpha,\alpha>=0$,
i.e. $\alpha=p^{+}b_{1}$. If $m<n-1$ and $<\alpha,b_{1}>=0$,
then $<\alpha,\alpha>\geq 0$.

Remark 9. Let us consider the enlargement of this subclass by the
addition of the terms  (3.26) to the Lagrangian. The modification
of the exact solution (3.59) for each $\sigma=2,\ldots,m$ is described
in the Remark 6. Let us take into account the additional components,
induced by isotropic vectors collinear to $b_{1}$ .  It is not
difficult to show that in this case (for $\sigma=1$) the additional
terms (3.26) leads to the following modification of the function
$f(t)$
\begin{eqnarray}
f(t)&&=\exp\{\frac{a^{(1)}}{2(p^{-})^{2}}\exp[p^{-}t+q^{-}]+
\sum_{\alpha=1}^{m(1)}\frac{a^{(1\alpha)}}{2b_{(1\alpha)}(p^{-})^{2}}
\exp[b_{(1\alpha)}(p^{-}t+q^{-})]\},\ p^{-}\neq 0,\nonumber \\
&&=\exp\{(a^{(1)}\exp[q^{-}]+\sum_{\alpha=1}^{m(1)}a^{(1\alpha)}
\exp[b_{(1\alpha)}q^{-}])\frac{t^{2}}{4}\},\ p^{-}=0.
\end{eqnarray}
In (3.56) and (3.61) the additional terms appear
\begin{equation}
\sum_{\alpha=1}^{m(1)}a^{(1\alpha)}\exp[b_{(1\alpha)}q^{-}].
\end{equation}
These are all modifications in this case.

\section{Reduction of pseudo-Euclidean Toda-like system
to  Euclidean one}
\par
Now we consider the case, when the set of vectors $b_{1},\ldots,b_{m}$ is
not orthogonal. It is easily shown that eqs. of motion of our system with the
Lagrangian
\begin{equation}
L=\frac{1}{2}<\dot{x},\dot{x}>-\sum_{\alpha=1}^{m}a^{(\alpha)}
\exp[<b_{\alpha},x>].
\end{equation}
for the new variables
\begin{eqnarray}
&&p=\dot{x}\in R^{n},\\
&&l_{\alpha}=a^{(\alpha)}\exp[<b_{\alpha},x>]
\end{eqnarray}
have the following form
\begin{eqnarray}
&&\dot{p}=-\sum_{\alpha=1}^{m}l_{\alpha}b_{\alpha},\\
&&\dot{l}_{\alpha}=l_{\alpha}<b_{\alpha},p>.
\end{eqnarray}
Note that this representation is valid for non-degenerate bilinear form
$<.,.>$ with arbitrary signature.

Let us consider a simple complex Lie algebra $G$. Let $H$ be a Cartan
subalgebra, and $h_{i},e_{\omega_{\gamma}}$ be a Weyl-Cartan basis in
$G$ [3]. We denote by $h_{1},\ldots,h_{n}$ some basis in $H$ and by
$\omega_{1},\ldots,\omega_{N}$ the set of roots ($\omega_{\gamma}\in H,\
\gamma=1,\ldots,N$). If the roots $\omega_{1},\ldots,\omega_{m}$ are
admissible, then we have [2]
\begin{eqnarray}
&&[h,e_{\omega_{\alpha}}]=(\omega_{\alpha},h)e_{\omega_{\alpha}},\ \
h\in H\\
&&[e_{\omega_{\alpha}},e_{-\omega_{\beta}}]=
\delta_{\alpha \beta}\omega_{\alpha},\ \ \alpha,\beta=1,\ldots,m,
\end{eqnarray}
where we denote by $(.,.)$ the Killing-Cartan form. Let us define in the
algebra $G$ the vectors ($L-A$ pair) [2]
\begin{eqnarray}
&&L(t)=\sum_{\alpha=1}^{m}f_{\alpha}(t)e_{-\omega_{\alpha}}+
C\sum_{i=1}^{n}h^{i}(t)h_{i}+C^{2}\sum_{\alpha=1}^{m}e_{\omega_{\alpha}},\\
&&A(t)=-\frac{1}{C}\sum_{\alpha=1}^{m}f_{\alpha}(t)e_{-\omega_{\alpha}},
\end{eqnarray}
where $C$ is arbitrary constant. Using (4.6-4.7), it can be easily checked
that eq.
\begin{equation}
\dot{L}(t)=[L(t),A(t)]
\end{equation}
is equivalent to the following set of eqs. for the variables
$f_{\alpha}(t),h^{i}(t)$
\begin{eqnarray}
&&\dot{h}=-\sum_{\alpha=1}^{m}f_{\alpha}\omega_{\alpha},\\
&&\dot{f}_{\alpha}=f_{\alpha}(\omega_{\alpha},h),
\end{eqnarray}
where we denoted $h=h^{1}(t)h_{1}+\ldots+h^{n}(t)h_{n},\ h\in H$.

Consider the real linear subspace of dimension $n$ $H^{\prime}\in H$ such
that the Killing-Cartan form $(.,.)$ on $H^{\prime}$ is a real non-degenerate
bilinear form with the signature $(-,+,\ldots,+)$, i.e. $<.,.>$. It is
evident, that the sets of eqs. (4.4-4.5) and (4.11-4.12) are identical, if
$h,\omega_{1},\ldots,\omega_{m}\in H^{\prime}$. Thus, if the set of vectors
$b_{1},\ldots,b_{m}\in R^{n}$ equipped with the bilinear form $<.,.>$ may be
identified with a set of admissible roots $\omega_{1}\ldots,\omega_{m}\in
H^{\prime}$, then pseudo-Euclidean Toda-like system with the Lagrangian (4.1)
possesses the Lax representation. If such identification is possible, then the
system is called to be connected with the simple complex Lie algebra.

Proposition 3.
Let a  pseudo-Euclidean  Toda-like  system  is connected with a
simple complex Lie algebra.  Then it is reducible to an Euclidean  Toda-like
system for a part of coordinates.

Proof.
We get in an  arbitrary  orthonormal basis
$e_{1}^{\prime},\ldots,e_{n}^{\prime}$
\begin{equation}
L=\frac{1}{2}\sum_{i,j=1}^{n}\eta_{ij}\dot{X}^{i}\dot{X}^{j}-
\sum_{s=1}^{m}a^{(s)}\exp[\sum_{i=1}^{n}B_{i}^{s}X^{i}],
\end{equation}
where we denoted
\begin{equation}
B_{i}^{s}=\sum_{j=1}^{n}\eta_{ij}B_{s}^{j}.
\end{equation}
We remind, that
$b_{s}=B_{s}^{1}e_{1}^{\prime}+\ldots+B_{s}^{n}e_{n}^{\prime}$.

It is  known  [3]  that  the Killing-Cartan form
defined on the real linear span of roots of a simple (or
semi-simple) complex Lie algebra is positively definite. But we have the
indefinite bilinear form $<.,.>$. Then the first components of the
vectors $b_{1},\ldots,b_{m}$
must be zero in a suitably chosen orthonormal basis, i.e.  $B_{1}^{s}=0$
for $s=1,\ldots,m$.  Then, in this basis Lagrangian (4.1) has the form
\begin{equation}
L=\frac{1}{2}\sum_{i,j=1}^{n}\eta_{ij}\dot{X}^{i}\dot{X}^{j}-
\sum_{s=1}^{m}a^{(s)}\exp[\sum_{k=2}^{m}B_{k}^{s}X^{k}].
\end{equation}
Coordinate $X^{1}$ satisfies the eq.: $\ddot{X}^{1}=0$ . Eqs.
of motion  for $X^{2},\ldots,X^{n}$ are followed from the Euclidean
Toda-like  Lagrangian
\begin{equation}
L_{E}=\frac{1}{2}\sum_{k,l=2}^{n}\delta_{kl}\dot{X}^{k}\dot{X}^{l}-
\sum_{s=1}^{m}a^{(s)}\exp[\sum_{k=2}^{m}B_{k}^{s}X^{k}].
 \end{equation}
Thus, we obtained the reduction of a pseudo-Euclidean Toda-like
system to the Euclidean one.

Integrating the eqs. of an Euclidean  Toda-like  system  by
known methods [4,5,7], we obtain the class of exact solutions for
some nonorthogonal set of vectors $b_{1},\ldots,b_{m}$. Here we
consider this procedure for the simplest $2$-component case
($n\geq 3$), when Toda lattice is connected with Lie algebra $A_{2}$.

Suppose, that the vectors $b_{1}$ and $b_{2}$,  inducing by two
components in the Lagrangian
\begin{equation}
L=\frac{1}{2}<\dot{x},\dot{x}>-a^{(1)}\exp[<b_{1},x>]-a^{(2)}\exp[<b_{2},x>],
\end{equation}
satisfy the following conditions
\begin{equation}
<b_{1},b_{2}>=-\frac{1}{2}<b_{1},b_{1}>=-\frac{1}{2}<b_{2},b_{2}><0.
\end{equation}
Then, we have two space-like vectors with the same lengths. The
angle between them is equal to $120^{\circ}$. We denote
\begin{equation}
\sqrt{<b_{1},b_{1}>}=\sqrt{<b_{2},b_{2}>}=b.
\end{equation}

Let us define the orthonormal basis $e_{1}^{\prime},\ldots,e_{n}^{\prime}$
in $R^{n}$ by
\begin{eqnarray}
b_{1}=be_{2}^{\prime}, \\
b_{2}=b(-\frac{1}{2}e_{2}^{\prime}+\frac{\sqrt{3}}{2}e_{3}^{\prime}).
\end{eqnarray}
In this basis the Lagrangian (4.17) and corresponding energy
constraint have the form
\begin{eqnarray}
&&L=\frac{1}{2}\sum_{i,j=1}^{n}\eta_{ij}\dot{X}^{i}\dot{X}^{j}-
a^{(1)}\exp[bX^{2}]-
a^{(2)}\exp[b(-\frac{1}{2}X^{2}+\frac{\sqrt{3}}{2}X^{3})], \\
&&E_{0}=\frac{1}{2}\sum_{i,j=1}^{n}\eta_{ij}\dot{X}^{i}\dot{X}^{j}+
a^{(1)}\exp[bX^{2}]+
a^{(2)}\exp[b(-\frac{1}{2}X^{2}+\frac{\sqrt{3}}{2}X^{3})]
\end{eqnarray}
For the coordinates $X^{1},\ X^{4},\ldots,X^{n}$ we get the following eqs.
of motion:
\begin{equation}
\ddot{X}^{1}=\ddot{X}^{4}=\ldots=\ddot{X}^{n}.
\end{equation}
Therefore
\begin{equation}
X^{1}=p^{1}t+q^{1},\ X^{4}=p^{4}t+q^{4},\ldots,X^{n}=p^{n}t+q^{n},
\end{equation}
where $p^{1},p^{4},\ldots,p^{n},q^{1},q^{4},\ldots, q^{n}$ are arbitrary
integration constants. The eqs. of motion for the coordinates
$X^{2}$ and ${X^3}$
follow from the Lagrangian
\begin{equation}
L_{E}=\frac{1}{2}((\dot{X}^{2})^{2}+(\dot{X}^{3})^{2})-
a^{(1)}\exp[bX^{2}]-
a^{(2)}\exp[b(-\frac{1}{2}X^{2}+\frac{\sqrt{3}}{2}X^{3})].
\end{equation}
Let us introduce new coordinates $y^{1}$ and $y^{2}$ as
\begin{equation}
y_{1}=\frac{b}{2\sqrt{2}}X^{2},\ \
y_{2}=\frac{b}{2\sqrt{2}}X^{3}.
\end{equation}
We obtain the Lagrangian of the open Toda lattice connected
with the Lie algebra $A_{2}=SL(3,C)$
\begin{equation}
L_{T}=\frac{1}{2}((\dot{y}^{1})^{2}+(\dot{y}^{2})^{2})-
\epsilon g_{1}^{2}\exp[2\sqrt{2}y_{1}]-
\epsilon g_{2}^{2}\exp[-\sqrt{2}y_{1}+\sqrt{6}y_{2}],
\end{equation}
where we denoted
\begin{eqnarray}
&&b^{2}a^{(1)}/8=\epsilon g_{1}^{2},\ \
b^{2}a^{(2)}/8=\epsilon g_{2}^{2},\\
&&\epsilon={\rm sgn}[a^{(1)}]={\rm sgn}[a^{(2)}]=\pm 1.
\end{eqnarray}
To study the open Toda lattice  it is useful to add the additional
coordinate $y_{3}$:
\begin{equation}
L_{T}=\frac{1}{2}((\dot{y}^{1})^{2}+(\dot{y}^{2})^{2} +(\dot{y}^{3})^{2})-
\epsilon g_{1}^{2}\exp[2\sqrt{2}y_{1}]-
\epsilon g_{2}^{2}\exp[-\sqrt{2}y_{1}+\sqrt{6}y_{2}],
\end{equation}
After the orthogonal linear transformation
\begin{eqnarray}
&&q_{1}=\frac{1}{\sqrt{6}}(\sqrt{3}y_{1}+y_{2}+\sqrt{2}y_{3}),\nonumber\\
&&q_{2}=\frac{1}{\sqrt{6}}(-\sqrt{3}y_{1}+y_{2}+\sqrt{2}y_{3}),\\
&&q_{3}=-2y_{2}+\sqrt{2}y_{3}
\end{eqnarray}
the Lagrangian (4.31) take the well-known form [1,4-8]
\begin{equation}
L_{T}=\frac{1}{2}(\dot{q}_{1}^{2}+\dot{q}_{2}^{2}+\dot{q}_{3}^{2})-
\epsilon g_{1}^{2}\exp[2(q_{1}-q_{2})]-
\epsilon g_{2}^{2}\exp[2(q_{2}-q_{3})].
\end{equation}
In this representation the additional degree of  freedom corresponds to the
free motion of the center of mass
($\ddot{q}_{1}+\ddot{q}_{2}+\ddot{q}_{3}=0$).  The integrating of the eqs. of
motion for this system leads to the result [4,5,7] \begin{equation}
g_{1}^{2}\exp[2(q_{1}-q_{2})]=\frac{F_{+}}{F_{-}^{2}},\ \ \
g_{2}^{2}\exp[2(q_{2}-q_{3})]=\frac{F_{-}}{F_{+}^{2}},
\end{equation}
where
\begin{eqnarray}
F_{\pm}=&&\frac{4}{9A_{1}A_{2}(A_{1}+A_{2})}\{A_{1}\exp[\pm(A_{1}+2A_{2})t\pm
B_{1}]+\\
&&\epsilon(A_{1}+A_{2})\exp[\pm(A_{1}-A_{2})t\mp(B_{1}-B_{2})]+
A_{2}\exp[\mp(2A_{1}+A_{2})t\mp B_{2}]\}\nonumber.
\end{eqnarray}
The integration constants $B_{1}$, $B_{2}$ are arbitrary and $A_{1}$,
$A_{2}$ satisfy the condition: $A_{1}A_{2}>0$. For the energy of the
system with Lagrangian (4.24) we have
\begin{equation}
\frac{1}{2}(\dot{q}_{1}^{2}+\dot{q}_{2}^{2}+\dot{q}_{3}^{2})+
\epsilon g_{1}^{2}\exp[2(q_{1}-q_{2})]+
\epsilon g_{2}^{2}\exp[2(q_{2}-q_{3})]=
\frac {3}{4}(A_{1}^{2}+A_{1}A_{2}+A_{2}^{2}).
\end{equation}
Doing the inverse linear transformation
\begin{eqnarray}
&&y_{1}=\frac{1}{\sqrt{2}}[q_{1}-q_{2}],\nonumber\\
&&y_{2}=\frac{1}{\sqrt{6}}([q_{1}-q_{2}]+2[q_{2}-q_{3}]),\\
&&y_{3}=\frac{1}{\sqrt{3}}(q_{1}+q_{2}+q_{3}),\nonumber
\end{eqnarray}
for the system with Lagrangian (4.22) we get the solution
\begin{eqnarray}
&&X^{2}=\frac{1}{b}\ln[\frac{8}{b^{2}|a^{(1)}|}\frac{F_{+}}{F_{-}^{2}}],\\
&&X^{3}=\frac{\sqrt{3}}{b}
\ln[\frac{8}{b^{2}(|a^{(1)}|(a^{(2)})^{2})^{1/3}}
\frac{1}{F_{+}}],
\end{eqnarray}
and the following energy constraint
\begin{equation}
E_{0}=-\frac{1}{2}(p^{1})^{2}+\frac{1}{2}(p^{4})^{2}+
\ldots+\frac{1}{2}(p^{n})^{2}+
\frac{6}{b^{2}}(A_{1}^{2}+A_{1}A_{2}+A_{2}^{2}).
\end{equation}

To present the scale factors in the Kasner-like form let us introduce the
Kasner-like parameters
\begin{eqnarray}
&&\alpha^{i}=t_{1}^{i}p^{1}+t_{4}^{i}p^{4}+\ldots+t_{n}^{i}p^{n},\\
&&\beta^{i}=t_{1}^{i}q^{1}+t_{4}^{i}q^{4}+\ldots+t_{n}^{i}q^{n},
\end{eqnarray}
where components $t_{k}^{i}$ are determined by (3.7). In this case they
satisfy the relations
\begin{equation}
t_{2}^{i}=\frac{1}{b}b_{1}^{i},\ \
t_{3}^{i}=\frac{2}{\sqrt{3}}(\frac{1}{b}b_{2}^{i}+\frac{1}{2b}b_{1}^{i}).
\end{equation}
 From (3.6) we obtain the coordinates $x^{i}$ and
finally present the exact solution in the form \begin{equation}
\exp[x^{i}]=[\tilde{F}_{-}^{2}]^{-b_{1}^{i}/<b_{1},b_{1}>}
[\tilde{F}_{+}^{2}]^{-b_{2}^{i}/<b_{2},b_{2}>}\exp[\alpha^{i}t+\beta^{i}],
\end{equation}
where
\begin{eqnarray}
&&\tilde{F}_{-}=\frac{1}{8}b^{2}\{(a^{(1)})^{2}|a^{(2)}|\}^{\frac{1}{3}}F_{-},\\
&&\tilde{F}_{+}=\frac{1}{8}b^{2}\{(a^{(2)})^{2}|a^{(1)}|\}^{\frac{1}{3}}F_{+}.
\end{eqnarray}
The vectors $\alpha$ and $\beta$ defined by (3.23) satisfy
the relations
\begin{eqnarray}
&&<\alpha,\alpha>=2(E_{0}-\frac{6}{b^{2}}(A_{1}^{2}+A_{1}A_{2}+A_{2}^{2})),\\
&&<\alpha,b_{r}>=<\beta,b_{r}>=0,\ \ r=1,2.
\end{eqnarray}

Remark 10. If $n=3$, then $<\alpha,\alpha>\leq 0$ and
$<\beta,\beta>\leq 0$.

\section{Discussion}
\par

Let us consider some cosmological models corresponding  to  the
introduced in the Sect. III   integrable   subclasses   of
pseudo-Euclidean Toda-like systems. For this purpose in Table I we present
values of the bilinear form $<.,.>$ (see Sect. II) for the vectors
\begin{eqnarray} &&v_{i}\equiv v_{(i)}^{1}e_{1}+\ldots+v_{(i)}^{n}e_{n},\ \
\ v_{(i)}^{j}=-2\frac{\delta_{i}^{j}}{N_{i}},\\ &&u_{\alpha}\equiv
u_{(\alpha)}^{1}e_{1}+\ldots+u_{(\alpha)}^{n}e_{n},\ \ \
u_{\alpha}^{j}=h_{j}^{(\alpha)}+
\frac{1}{2-D}\sum_{i=1}^{n}N_{i}h_{i}^{(\alpha)},\\
&&u\equiv u^{1}e_{1}+\ldots+u^{n}e_{n},\ \ \
u^{j}=\frac{2}{2-D},
\end{eqnarray}
induced by curvature, perfect fluid and $\Lambda $-term correspondingly.

Within the subclass A  we are able to construct the model with one
Einstein space of
non-zero curvature. Let $(n-1)$  Einstein  spaces  are
Ricci-flat and one, for instance $M_{1}$ , have a non-zero Ricci tensor.
Then we put $b_{1}\equiv v_{1}$. To get the orthogonality with $b_{1}$
for at most $(n-1)$  available components of the perfect fluid
($b_{(\alpha+1)}\equiv u_{(\alpha)}$ for $\alpha\leq n-1$)
 we put: $h_{1}^{(\alpha)}=0 $ (see Table I).
Then, these components appeared to be  in the manifold $M_{1}$ in the
Zeldovich matter form (see Remark 1). The model of such a type were
integrated in [32]. In the same way the model with all Ricci-flat spaces
and $\Lambda$-term arises . In this case we put $b_{1}=u$.  The condition
of the orthogonality reads: $\sum_{i=1}^{n}h_{i}^{(\alpha)}N_{i}=0$ for
all $\alpha\leq n-1$. Then we get the negative values for the some
$h_{i}^{(\alpha)}$. It means that for such perfect fluids
$p>\rho$ in some  spaces (see (2.8)).

The vectors $v_{i}$ and $u$
induced by curvature and $\Lambda$-term correspondingly
are time-like, therefore subclasses B and  C correspond to  the
Ricci-flat models without $\Lambda$-term  for some
multicomponent perfect fluid source. These vectors can not be roots of any
simple complex Lie algebra.  Therefore,   the
models with more than one non-zero curvature space and the models with
curvature and $\Lambda$-term are not trivially reducible to the Euclidean
Toda lattices. Some possibilities of integration of these models were
studied in [27,29].

In conclusion we discuss the existence of the Euclidean wormholes [37-40]
within the class of the obtained exact solutions. We consider the simple model
within subclass A with the manifold $R\times M_{1}\times M_{2}$, when
$M_{1}$ has a nonzero Ricci tensor with $\lambda_{1}>0$ (see 2.3) and $M_{2}$
is Ricci-flat. The integrable model arises in the presence of the perfect
fluid in the Zeldovich matter form for the space $M_{1}$. It means
$h_{1}=0$  and the other parameter in the equation of state for $M_{2}$
(see 2.8)) may be arbitrary positive constant $h$. If we demand the
positiveness of the mass-energy density for the perfect fluid ($A>0$), then
from (3.22) we get for the scales factors of the  $M_{1}$ and $M_{2}$
\begin{eqnarray}
&&\exp[x^{1}]=\{F_{1}^{2}(t-t_{01})\}^{-\frac{1}{2(N_{1}-1)}}
\{F_{2}^{2}(t-t_{02})\}^{\frac{1}{h(N_{1}-1)}},\\
&&\exp[x^{2}]=\{F_{2}^{2}(t-t_{02})\}^{-\frac{1}{hN_{2}}},
\end{eqnarray}
where
\begin{eqnarray}
&&F_{1}(t-t_{01})=\sqrt{\frac{1}{2}\lambda_{1}N_{1}/|E_{1}|}
\cosh[\sqrt{2|E_{1}|(N_{1}-1)/N_{1}}(t-t_{01})],\\
&&F_{2}(t-t_{01})=\sqrt{\kappa^{2}A/E_{2}}
\cosh[h\sqrt{\frac{1}{2}(N_{1}-1)N_{2}|E_{2}/(N_{1}+N_{2}-1)}(t-t_{02})].
\end{eqnarray}
In this case $E_{1}<0$ and $E_{2}>0$. The energy constraint (3.24)
leads to the condition: $-E_{1}=E_{2}\equiv E$.

We may suppose that $M_{1}$ is 3-dimensional sphere $S^{3}$ and $M_{2}$
is $d$-dimensional torus $T^{d}$. Then formulas (5.4-5.7) present
the multidimensional generalization of closed Friedmann model. This model may
be relevant in the theory of the Early Universe, because the Zeldovich matter
equation of state: $p=\rho$ is valid on the earlier stage of its evolution
[35].

To prove the existence of the Euclidean wormholes we use the transformation
$t\rightarrow it$ . Then for the case $t_{01}=t_{02}=0 $ we obtain
\begin{eqnarray}
&&\exp[x^{1}]=\{\frac{\kappa^{2}A}{E}\cos^{2}[\sqrt{\frac{Ed}{d+2}}ht]\}^{1/(2h)}
\{\frac{3\lambda_{1}}{2E}\cos^{2}[\sqrt{\frac{4E}{3}}t]\}^{-1/4},\\
&&\exp[x^{2}]=\{\frac{\kappa^{2}A}{E}\cos^{2}[\sqrt{\frac{Ed}{d+2}}ht]\}^{-1/(hd)}.
\end{eqnarray}
It is easy to see that when $\frac{d}{d+2}h^{2}>\frac{4}{3}$ one has wormhole
with respect to the internal space $T^{d}$. The case
$\frac{d}{d+2}h^{2}<\frac{4}{3}$ corresponds to the wormhole for the
external space $S^{3}$. Note, that for $h=2$ and $d=1$ the wormhole for the
internal space is accompanied by the static external space. It is not
difficult to show that wormhole with respect to the whole space for this
model arises in the presence of the additional component in the form of
minimally coupled scalar field.

\begin{center}
{\bf Acknowledgments}
\end{center}
\par
The authors are grateful to Prof. D.-E.Liebscher for useful discussions.
\par
This work was supported in part by the Russian Ministry of Science.

\pagebreak

\pagebreak

$
\begin{array}{lcccc}
& & & & \\
\hline \hline
&\ \  & v_{j}  & u_{\beta} & u\\ \hline
&\ \ & & & \\
v_{i}&\ \  & 4(\frac{\delta_{ij}}{N_{i}}-1) &-2h_{i}^{(\beta)} &-4\\
&\ \ & & & \\
u_{\alpha} &\ \   &-2h_{j}^{(\alpha)}&\sum_{i=1}^{n}h_{i}^{(\alpha)}
h_{i}^{(\beta)} N_{i}+
&\frac{2}{2-D}\sum_{i=1}^{n}h_{i}^{(\alpha)}N_{i}\\
&\ \  & &\frac{1}{2-D}[\sum_{i=1}^{n}h_{i}^{(\alpha)} N_{i}]
[\sum_{j=1}^{n}h_{j}^{(\beta)} N_{j}] & \\
&\ \ & & & \\
u &\ \  & -4 &\frac{2}{2-D}\sum_{i=1}^{n}h_{i}^{(\beta)}N_{i}
&-4\frac{D-1}{D-2}\\
& & & &\\
\hline \hline
& & & &
\end{array}
$
\begin{center}
{TABLE I}
\end{center}

\end{document}